\documentclass[12pt]{article}

\usepackage{amsmath}
\usepackage{amssymb}
\usepackage{epsfig}
\usepackage{graphicx}
\usepackage{cite}
\usepackage{bbm}

\allowdisplaybreaks[1]

\makeatletter
\renewcommand{\fnum@table}{\textbf{\tablename~\thetable}}
\renewcommand{\fnum@figure}{\textbf{\figurename~\thefigure}}
\makeatother

\newcounter{myenumi}

\renewcommand{\themyenumi}{\roman{myenumi}}
{\end{list}}

\newlength{\myem}
\newcounter{mysubequation}[equation]

\makeatletter
\renewcommand{\section}{\@startsection{section}{1}{0em}{-\baselineskip}%
{\baselineskip}{\normalfont\large\bfseries}}
\renewcommand{\subsection}%
{\@startsection{subsection}{2}{0em}{-0.7\baselineskip}%
{0.7\baselineskip}{\normalfont\bfseries}}
\makeatother

\newcommand{\bi}{\begin{itemize}}
\newcommand{\ei}{\end{itemize}}

\newcommand{\be}{\begin{equation}}
\newcommand{\ee}{\end{equation}}
\newcommand{\bea}{\begin{eqnarray}}
\newcommand{\eea}{\end{eqnarray}}

\newcommand{\ie}{{\it i.e.}}

\newcommand{\eg}{{\it e.g.}}

\newcommand{\cf}{{\it cf.}}

\newcommand{\eq}{Eq.}
\newcommand{\eqs}{Eqs.}

\newcommand{\Fig}{Fig.}

\newcommand{\Figs}{Figs.}
\newcommand{\Ref}{Ref.}
\newcommand{\Refs}{Refs.}
\newcommand{\Sec}{Sec.}

\newcommand{\App}{Appendix}

\begin{document}

\begin{titlepage}

\renewcommand{\thefootnote}{\alph{footnote}}

\renewcommand{\thefootnote}{\fnsymbol{footnote}}
\setcounter{footnote}{-1}

{\begin{center} {\large\bf Indirect Detection of Kaluza--Klein Dark
Matter from Latticized Universal Dimensions } \end{center}}
\renewcommand{\thefootnote}{\alph{footnote}}

\vspace*{.8cm}
\vspace*{.3cm}
{\begin{center} {\large{\sc
 		Tomas~H{\"a}llgren\footnote[1]{\makebox[1.cm]{Email:}
                tomashal@kth.se} and
                Tommy~Ohlsson\footnote[2]{\makebox[1.cm]{Email:}
                tommy@theophys.kth.se}~
                }}
\end{center}}
\vspace*{0cm}
{\it
\begin{center}

\footnotemark[1]${}^,$\footnotemark[2]%
Department of Theoretical Physics, School of
Engineering Sciences, Royal Institute of Technology (KTH) -- AlbaNova
University Center,\\
Roslagstullsbacken 11, 106~91~~Stockholm, Sweden

\end{center}}

\vspace*{1.5cm}

{\Large \bf
\begin{center} Abstract \end{center}}

We consider Kaluza--Klein dark matter from latticized universal
dimensions. We motivate and investigate two different lattice models,
where the models differ in the choice of boundary conditions. The
models reproduce relevant features of the continuum model for
Kaluza--Klein dark matter. For the model with simple boundary
conditions, this is the case even for a model with only a few lattice
sites. We study the effects of the latticization on the differential
flux of positrons from Kaluza--Klein dark matter annihilation in the
galactic halo. We find that for different choices of the
compactification radius, the differential positron flux rapidly
converges to the continuum model results as a function of the number
of lattice sites. In addition, we consider the prospects for upcoming
space-based experiments such as PAMELA and AMS-02 to probe the
latticization effect.

\vspace*{.5cm}

\end{titlepage}

\newpage

\renewcommand{\thefootnote}{\arabic{footnote}}
\setcounter{footnote}{0}

\section{Introduction}

Today, cosmological measurements by, \eg, the Wilkinson Microwave
Anisotropy Probe (WMAP) \cite{Spergel:2003cb,Spergel:2006hy}, the
Two-Degree Field Galaxy Redshift Survey (2dFGRS)
\cite{Percival:2002gq}, and the Sloan Digital Sky Survey (SDSS)
\cite{Tegmark:2003ud} indicate that the energy and matter in the
Universe could be distributed such that approximately 73~\% is ``dark
energy'' (maybe due to the cosmological constant $\Lambda$ in
Einstein's field equations) and approximately 23~\% is ``dark
matter'', which leaves around 4~\% as ordinary luminous matter in
terms of baryons such as protons and neutrons, \ie, baryonic matter.

Further evidence for the presence of dark matter comes from, \eg,
galactic rotational curves \cite{Persic:1995ru,Sofue:2000jx} and
studies of the observed redshift-luminosity relation of type Ia
supernovae \cite{Knop:2003iy,Riess:2004nr}. All these data
consistently point towards approximately 23~\% dark matter. However,
the nature of the dark matter is still unknown. Nevertheless, several
dark matter candidates exist. Ordinary matter could make up a fraction
of the dark matter in the form of massive compact halo objects
(MACHOs) \cite{Paczynski:1985jf,Shaham:1979jf}. This could be, \eg,
faint stars or stellar remnants such as black holes. However, the
MACHOs can not make up all of the dark matter.

One of the most plausible candidates for dark matter is weakly
interacting massive particles (WIMPs)
\cite{Lee:1977ua,Gunn:1978gr}. These are hypothetical non-baryonic
particles with masses in the GeV-TeV range. They are electrically
neutral and assumed to carry a conserved quantum number to ensure
their stability. Within this category, the prime candidate is the
lightest supersymmetric particle -- typically the neutralino
\cite{Ellis:1983ew}. It is electrically neutral and stable due to the
conservation of R-parity.

Recently, an interesting alternative dark matter candidate has been
intensively studied in the literature, which is so-called
Kaluza--Klein dark matter \cite{Servant:2002aq,Cheng:2002ej}. In
models with universal extra dimensions
\cite{Appelquist:2000nn,Cheng:2002iz} (for the first model with TeV
sized extra dimensions, see \cite{Antoniadis:1990ev}), in which all
standard model particles can move, the first excited state or mode of
the hypercharge gauge boson is an excellent candidate for dark
matter. The prospects for both direct and indirect detection are
generally good. In particular, for indirect detection, the prospects
are better than for supersymmetric particles. This is because the
Kaluza--Klein gauge boson is a vector particle, whereas the neutralino
is a Majorana particle, and thus, has a helicity suppressed
annihilation cross-section. Therefore, one could obtain an excess of
positrons from Kaluza--Klein dark matter annihilation in the galactic
halo \cite{Hooper:2004xn,Hooper:2004bq} (for neutralino dark matter
detection with positrons, see also
\cite{Kamionkowski:1990ty,Baltz:1998xv}). This positron flux could,
\eg, be detected in balloon-borne experiments such as HEAT
\cite{Coutu:2001jy} or future space-based experiments such as PAMELA
\cite{Boezio:2004jx} and AMS-02 \cite{Barao:2004ik,Bosio:2004dh}.

In general, a major disadvantage for models with extra spatial
dimensions is that they have coupling constants with negative mass
dimension and are therefore not renormalizable. Hence, these models
can only be considered as effective models, valid up to some cutoff
energy scale, where some more fundamental theory is expected to
describe physics. The cutoff procedure leads after dimensional
reduction to an effective four dimensional description, but without
the full higher-dimensional gauge invariance. One possibility to
remedy these problems is offered by deconstructed
\cite{Arkani-Hamed:2001ca} or latticized \cite{Hill:2000mu} extra
dimensions.\footnote{Note that what is referred to as deconstructed or
latticized extra dimensions may differ in the literature. Here we
follow the classification of \Ref~\cite{Oliver:2003cc}.}

We will in this paper consider latticized extra dimensions. In these
type of models, the fundamental high-energy theory is four dimensional
with a replicated gauge symmetry of the form $G \times G \times \ldots
\times G$, where $G$ is a gauge group. In addition, it contains a set
of scalar fields, so-called link fields, which ``link'' the gauge
groups. If the potential of the scalar fields is chosen in an
appropriate way, then one can generate the physics of an extra
dimension in the infra-red regime. This happens when the link fields
acquire vacuum expectation values and spontaneously break the original
symmetry down to a diagonal subgroup of the symmetry. The resulting
low-energy model is a model with an extra dimension, where the extra
dimension has been put on a lattice. Note that the model with
deconstructed dimensions is structurally similar but replaces the link
scalar fields with fermion condensates.

In this way, models with deconstructed or latticized extra dimensions
can have similar benefits as models with continuum extra dimensions,
but since they in contrast with the latter preserves manifest gauge
invariance and sometimes are renormalizable quantum field theories,
they have a better defined short distance behavior. This could be
relevant to models for Kaluza--Klein dark matter, where corrections
from physics above the ultra-violet cutoff energy scale could be
crucial when determining the nature of the lightest Kaluza--Klein
particle \cite{Cheng:2002iz}. Radiative corrections to Kaluza--Klein
masses have been calculated for latticized abelian
\cite{Falkowski:2003iy} and non-abelian \cite{Kunszt:2004ps} gauge
theories compactified on $M^{4}\times S^{1}$. So far, these
calculations have not been applied to the case of latticized universal
extra dimensions, where there should be additional contributions to
Kaluza--Klein masses from the orbifold fixed points. In this paper, we
will not consider radiative corrections, but instead consider the
prospects for observing the latticized model already at tree-level.

More specifically, we will consider two lattice models for a universal
extra dimension, where the models differ in the choice of boundary
conditions. The models are designed to mimic in the infra-red regime
the continuum theory for Kaluza--Klein dark matter. In the
ultra-violet regime, the higher-dimensional theory is replaced with
completely four dimensional dynamics. We study the effects of the
latticization and whether one could probe these effects in future
experiments such as PAMELA and AMS-02.

The paper is organized as follows. In \Sec~\ref{sec:LUD}, we describe
the model for a latticized universal dimension with simple boundary
conditions. Next, in \Sec~\ref{sec:indet}, we study the flux of
positrons coming from Kaluza--Klein dark matter annihilations in the
galactic halo and the prospects for probing the latticization in
upcoming space-based experiments. Then, in \Sec~\ref{sec:MBC}, we
consider a lattice model with modified boundary conditions. Finally,
in \Sec~\ref{sec:S&C}, we summarize our results and present our
conclusions.

\section{Latticized Universal Dimensions}
\label{sec:LUD}

In this section, we will follow closely the results of
\Refs~\cite{Cheng:2001vd,Oliver:2003cc}.

The model we will consider is a field theory in four dimensions with a
product gauge group $G=\Pi_{j=0}^{N}SU(3)_{j}\times SU(2)_{j} \times
U(1)_{j}$. The model contains fermions and gauge bosons as well as a
set of scalar link fields $Q_{j,j+1},\Phi_{j,j+1}$, and
$\phi_{j,j+1}$, where $j=0,1,\ldots, N-1$. The link fields transform
as bifundamentals under adjacent gauge groups.

When the link fields acquire vacuum expectation values (VEVs), \ie,
$\langle Q_{j,j+1}\rangle=v_{3}\mathbbm{1}_{3}$, $\langle
\Phi_{j,j+1}\rangle=v_{2}\mathbbm{1}_{2}$, and $\langle
\phi_{j,j+1}\rangle=v_{1}/\sqrt{2}$, where $\mathbbm{1}_n$ is the $n
\times n$ identity matrix, the product gauge group is
spontaneously broken down to the diagonal subgroup $SU(3)\times
SU(2)\times U(1)$, which we identify as the standard model (SM) gauge
group. One can arrange the parameters of the scalar potential in such
a way that the link fields become non-linear sigma model fields in the
low-energy effective theory. The low-energy effective theory can then
be identified with a transverse lattice gauge theory
\cite{Bardeen:1979xx}, where only the extra dimension has been
latticized. Since we are interested in a model that mimics the
continuum theory for Kaluza--Klein dark matter, we will consider a
latticized version of an $S^{1}/\mathbb{Z}_{2}$ orbifold. This
topology has in the continuum theory the advantages of generating
chiral zeroth modes and of removing unwanted scalar degrees of
freedom\footnote{The zeroth mode of the fifth component of the
higher-dimensional gauge field, $A_{5}^{(0)}$.}, in agreement with
observations. Also, the orbifold topology imply a discrete symmetry --
Kaluza--Klein parity, which ensures the stability of the lightest
Kaluza--Klein particle, and thus, making it a viable dark matter
candidate. We will find that the lattice model considered in this
section reproduces the essential features of the continuum
$S^{1}/\mathbb{Z}_{2}$ orbifold.\footnote{We will not explicitly show
the absence of $A_{5}^{(0)}$. For a discussion of this, see for
example \Refs~\cite{Hill:2000mu} and \cite{Cheng:2001vd}.}

The action for this theory can be split as $S=S_{{\rm gauge}}+S_{{\rm
fermion}}+S_{{\rm Higgs}}$. In what follows, we will neglect
electroweak symmetry breaking effects and so we will not further
consider the contribution from the Higgs sector.

\subsection{Gauge Sector}

We will take the action of the gauge sector to be \cite{Cheng:2001vd}
\begin{eqnarray}
S_{{\rm gauge}}& = &\int {\rm d}^{4}x \bigg\lbrack \sum_{j=0}^{N-1}
\left( |D_{\mu}\phi_{j,j+1}|^{2}+{\rm
Tr}|D_{\mu}\Phi_{j,j+1}|^{2}+{\rm Tr}|D_{\mu}Q_{j,j+1}|^{2} \right)
\nonumber\\ &&- V(\phi,\Phi,Q)+ \sum_{j=0}^{N} \left(
-\frac{1}{4}F_{j\mu\nu}F^{\mu\nu}_{j}-\frac{1}{4}F_{j\mu\nu}^{a}F^{\mu\nu
a}_{j} -\frac{1}{4}F_{j\mu\nu}^{b}F^{\mu\nu b}_{j}
\right)\bigg\rbrack,
\end{eqnarray}
where $a=1,2,3$ and $b=1,2,\ldots,8$. Here $V(\phi,\Phi,Q)$ is a
suitably chosen scalar potential. The $Q_{j,j+1}$ fields transform
under $SU(3)_{j}\times SU(3)_{j+1}$ as $(\bar{{\bf 3}},{\bf 3})$,
$\Phi_{j,j+1}$ transform under $SU(2)_{j}\times SU(2)_{j+1}$ as
$({\bar{\bf 2}},{\bf 2})$, and $\phi_{j,j+1}$ are charged under
$U(1)_{j}\times U(1)_{j+1}$ as $(-Y_{\phi}, Y_{\phi})$. We will set
$Y_{\phi}=1/3$.

The covariant derivatives act on the link fields as
\begin{align}
D_{\mu}\phi_{j,j+1} &= \left(\partial_{\mu}+{\rm
i}\tilde{g}_{Y}\frac{Y_{\phi}}{2}A_{j\mu}-{\rm
i}\tilde{g}_{Y}\frac{Y_{\phi}}{2}A_{(j+1)\mu}\right)\phi_{j,j+1}, \\
D_{\mu}\Phi_{j,j+1} &= (\partial_{\mu}+{\rm
i}\tilde{g}A_{j\mu}^{a}T_{j}^{a}-{\rm
i}\tilde{g}A_{(j+1)\mu}^{a}T_{j+1}^{a})\Phi_{j,j+1}, \\
D_{\mu}Q_{j,j+1} &= (\partial_{\mu}+{\rm
i}\hat{g}A_{j\mu}^{b}T_{j}^{b}-{\rm
i}\hat{g}A_{(j+1)\mu}^{b}T_{j+1}^{b})Q_{j,j+1}.
\end{align}
where $T_{j}^{a}\,(a=1,2,3)$ are the generators of $SU(2)_{j}$,
$T_{j}^{b}\,(b=1,2,\ldots 8)$ are the generators of $SU(3)_{j}$, and
$\tilde{g}_{Y}$, $\tilde{g}$, and $\hat{g}$ are the coupling constants
of $U(1)_{j}$, $SU(2)_{j}$, and $SU(3)_{j}$, respectively. We have
assumed discrete translational invariance by setting
$\tilde{g}_{Yj}\equiv \tilde{g}_{Y}$, $\tilde{g}_{j}\equiv \tilde{g}$,
and $\hat{g}_{j}\equiv \hat{g}$, for all $j$. When the link fields
acquire VEVs, the kinetic terms for the link fields generate mass
matrices for the gauge bosons. Thus, for the $U(1)$ gauge fields, we
obtain the mass terms
\begin{equation}\label{eq:massmatrix1}
\mathcal{L}_{{\rm mass}} = \frac{1}{8} \tilde{g}_{Y}^{2} v_{1}^{2}
Y_{\phi}^{2} \sum_{j=0}^{N-1}(A_{j}-A_{j+1})^{2}.
\end{equation}
Similarly, for the other gauge fields, we obtain the same kind of mass
terms.

Explicitly, the $(N+1)\times (N+1)$ tridiagonal mass-squared matrix reads
\begin{equation}
M^{2}=\frac{1}{8}\tilde{g}_{Y}^{2}v_{1}^{2}Y_{\phi}^{2}\left(\begin{matrix}
1 & -1\\ -1 & 2 & -1 \\ & \ddots & \ddots & \ddots \\ & & -1 & 2 &
-1\\ & & & -1 & 1
\end{matrix}\right),
\end{equation}
which can be diagonalized by a change of basis
\begin{equation}
A_{j}=\sum_{n=0}^{N}a_{jn}\tilde{A}_{n},
\end{equation}
where
\begin{equation}\label{eq:ajn}
a_{jn}=\left\{ \begin{array}{ll} 
\sqrt{\frac{2}{N+1}}\cos\left(\frac{2j+1}{2}\gamma_{n}\right), & n\neq 0\\
\sqrt{\frac{1}{N+1}}, & n=0 \end{array} \right..
\end{equation}
Here $\gamma_{n}=n\pi/(N+1)$. In this basis, we now obtain
\begin{equation}\label{eq:gaugebosonmasses}
\mathcal{L}_{{\rm
mass}}=\frac{1}{2}\sum_{n=0}^{N}m_{n}^{2}\tilde{A_{n}}\tilde{A_{n}},
\end{equation}
where
\begin{equation}\label{eq:gbmass}
m_{n}^{2} = \tilde{g}_{Y}^{2} v_{1}^{2} Y_{\phi}^{2} \sin^{2}
\left[\frac{n\pi}{2(N+1)}\right].
\end{equation}
In the limit $n\ll N$, we find a linear Kaluza--Klein spectrum $m_{n}
\simeq n/R$ provided that we make the identification $\pi
\tilde{g}_{Y} v_{1} Y_{\phi} /[2(N+1)]=1/R$. It should be noted that
when referring to $R$, we refer to the radius of the extra dimension,
whereas some other authors with the same notation refer to the size of
the extra dimension (\ie, $\pi R$ for an $S^{1}/\mathbb{Z}_{2}$
orbifold).

\subsection{Fermionic Sector}

\subsubsection{Fermions}\label{sec:fermions}

Next, we include a set of fermions $L^{\alpha}_{j}=(\nu^{\alpha}_{j},
e^{\alpha}_{j})^{T}$ and $E^{\alpha}_{j}$, for $j=0,1,\ldots,N$ and
$\alpha=e,\mu,\tau$ (from now on generational indices will be
suppressed). Here $L_{j}$ transforms as ${\bf 2}$ under $SU(2)_{j}$
and as a singlet under $SU(2)_{i}$, for $i\neq j$. Furthermore,
$L_{j}$ is charged under $U(1)_{j}$ as $Y_{d}=-1$. The field $E_{j}$
is a singlet under all $SU(2)$ groups and is charged under $U(1)_{j}$
as $Y_{s}=-2$. Both $L_{j}$ and $E_{j}$ transform trivially under all
$SU(3)$ groups.

In \Ref~\cite{Oliver:2003cc}, the latticized action for the quark
sector was considered. We construct analogously the action for leptons
as $S_{{\rm fermion}}=S_{d}+S_{s}$, where $S_{d}$ refers to the part
containing the $SU(2)$ doublet fields and $S_{s}$ contains the $SU(2)$
singlet fields. Here $S_{d}$ is given by
\begin{align}\label{eq:Sd}
S_{d} &= \int {\rm d}^{4}x\,\left\{\sum_{j=0}^{N}\bar{L}_{j}{\rm
i}\gamma^{\mu}D_{\mu}L_{j} - \sum_{j=0}^{N} \left[ M_{f}{\bar L}_{jL}
\left( \frac{\Phi^{\dagger}_{j,j+1}}{v_{2}}
\frac{\phi_{j,j+1}^{3}}{(v_{1}/\sqrt{2})^{3}} L_{j+1,R} - L_{jR}
\right) + {\rm h.c.} \right]\right\}, \nonumber\\
\end{align}
where $M_{f}$ is a mass parameter which will be used when matching to
the continuum model. The covariant derivative is given by
$D_{\mu}L_{j}=\left(\partial_{\mu}-{\rm
i}\tilde{g}A^{a}_{j\mu}T_{j}^{a}-{\rm
i}\tilde{g}_{Y}\frac{Y_{d}}{2}A_{j\mu}\right)L_{j}$. If the index of a field
is out of bounds, then it is implicit that the field is zero. For
the $SU(2)$ singlet fields, we have
\begin{equation}\label{eq:Ss}
S_{s} = \int {\rm d}^{4}x\, \left\{ \sum_{j=0}^{N}{\bar E}_{j}{\rm
i}\gamma^{\mu}D_{\mu}E_{j} + \sum_{j=0}^{N} \left[ M_{f}{\bar E}_{jR}
\left(\frac{\phi_{j,j+1}^{6}}{(v_{1}/\sqrt{2})^{6}}E_{j+1,L} - E_{jL}
\right) + {\rm h.c.} \right]\right\},
\end{equation}
where the covariant derivate is given by
$D_{\mu}E_{j}=\left(\partial_{\mu}-{\rm i}
\tilde{g}_{Y}\frac{Y_{s}}{2}A_{j\mu}\right)E_{j}$. Note that the signs for
the mass terms in \eqs~(\ref{eq:Sd}) and (\ref{eq:Ss}) are
different. This is a characteristic feature of theories with
continuous universal extra dimensions after dimensional reduction
\cite{Appelquist:2000nn}. Here, as in \Ref~\cite{Cheng:2001nh}, it has
been obtained by choosing different signs for the Wilson terms when
deriving \eqs~(\ref{eq:Sd}) and (\ref{eq:Ss}). The signs for the mass
terms are also in agreement with the corresponding results for the
quark sector in \Ref~\cite{Oliver:2003cc}.

In order to obtain chiral zeroth modes, we take the doublet fields to
satisfy $L_{0R}=0$ and the singlet fields to satisfy $E_{0L}=0$. When
the link fields acquire universal VEVs, we obtain mass matrices for
the fermion fields. For example, in the doublet sector, we have
\begin{equation}
\mathcal{L}_{{\rm
mass}}=-M_{f} \left[\bar{L}_{0L}L_{1R}+\sum_{j=1}^{N}{\bar
L}_{jL}\left(L_{j+1,R}-L_{jR}\right)+{\rm h.c.}\right].
\end{equation}
The corresponding mass matrix is diagonalized by a change of basis
\begin{equation}
L_{jL}=\sum_{n=0}^{N}a_{jn}\tilde{L}_{nL}\,,\quad\quad
L_{jR}=\sum_{n=1}^{N}b_{jn}\tilde{L}_{nR}\,,
\end{equation}
where the $a_{jn}$'s are given in \eq~(\ref{eq:ajn}) and
\begin{equation}\label{eq:bjn}
b_{jn}=\sqrt{\frac{2}{N+1}}\sin\left(j\gamma_{n}\right).
\end{equation}
Thus, we obtain for the left-handed fields the masses
\begin{equation}\label{eq:fermionmass}
m_n^2 = 4 M_f^{2}\sin^2 \left[\frac{n\pi}{2(N+1)}\right],
\end{equation}
where $n=0,1,\ldots, N$ and for the right-handed fields the same form
for the masses, but now $n=1,2,\ldots, N$. Thus, there are no zeroth
modes for the right-handed doublet fields. For $n\ll N$, we find a
linear Kaluza--Klein spectrum if we make the identification $\pi
M_{f}/(N+1)=1/R$. Note that with this identification the masses in
\eq~(\ref{eq:fermionmass}) become identical to the masses in
\eq~(\ref{eq:gbmass}), after the corresponding identification of
parameters has been made for \eq~(\ref{eq:gbmass}). This is expected
and it motivates the use of the same notation in
\eqs~(\ref{eq:gbmass}) and (\ref{eq:fermionmass}).

Similarly, we obtain a mass matrix for the singlet sector, which is
diagonalized by a change of basis
\begin{equation}
E_{jL}=\sum_{n=1}^{N}b_{jn}\tilde{E}_{nL}\,,\quad\quad
E_{jR}=\sum_{n=0}^{N}a_{jn}\tilde{E}_{nR}.
\end{equation}
Note that $L_{jL}$ has the same expansion as $E_{jR}$, whereas
$L_{jR}$ has the same expansion as $E_{jL}$. Furthermore, note that we
obtain a right-handed zeroth mode and no left-handed zeroth modes for
the singlet fields. This is because we have interchanged the role of
the left- and right-handed fields in this case.

Note that, as in the continuum model, we obtain negative masses for
the singlet fields. However, this can be remedied by a change of basis
$\tilde{E}_{n}\rightarrow-\gamma_{5}\tilde{E}_{n}$, for
$n=1,2,\ldots,N$ \cite{Appelquist:2000nn}.

\subsubsection{Fermion Gauge Boson Couplings}\label{sec:ffA}

Now, we consider the couplings between fermions and the $U(1)$ gauge
bosons. For the $SU(2)$ doublet fields, we obtain from the fermionic
kinetic terms $\sum_{j=0}^{N}\bar{L}_{j}{\rm
i}\gamma^{\mu}D_{\mu}L_{j}$, the couplings
$\tilde{g}_{Y}\frac{Y_{d}}{2}
\sum_{j=0}^{N}\bar{L}_{j}\gamma^{\mu}A_{j\mu}L_{j}$. In the mass
eigenbasis, we have for the left-handed fields
\begin{equation}
\mathcal{L}_{ffA} = \tilde{g}_{Y}\frac{Y_{d}}{2}\sum_{n,m,l}
\sum_{j}a_{jn}a_{jm}a_{jl}
\bar{\tilde{L}}_{nL}\tilde{A}_{m\mu}\gamma^{\mu}\tilde{L}_{lL}.
\end{equation}
Using the orthogonality relations given in \App~\ref{app:orthogonal},
we have
\begin{equation}
\mathcal{L}_{ffA} = \tilde{g}_{Y}\frac{Y_{d}}{2\sqrt{N+1}}
\bar{\tilde{L}}_{1L}\tilde{A}_{1\mu}\gamma^{\mu}\tilde{L}_{0L} + {\rm h.c.}
 + \ldots,
\end{equation}
where the dots indicate terms that are not directly relevant to our
discussion. In the continuum model, the corresponding coupling is
given by \cite{Bergstrom:2004nr,Appelquist:2000nn}
\begin{equation}\label{eq:contd}
\mathcal{L}_{{\rm cont.}} =
\frac{1}{4}g_{Y}Y_{d}\bar{L}_{L}^{(1)}A_{\mu}^{(1)}
\gamma^{\mu}(1-\gamma_{5})L^{(0)} + {\rm h.c.}.
\end{equation}
There is an additional factor $1/2$ in \eq~(\ref{eq:contd}) which is
not present in \Ref~\cite{Bergstrom:2004nr}. This is simply due to the
fact that we have chosen a different convention for the hypercharge
assignment. If we identify $\tilde{g}_{Y}/\sqrt{N+1}=g_{Y}$, then we
observe that we have the same coupling as for the continuum case.

For the singlet fields, we obtain from the kinetic terms
$\sum_{j=0}^{N}\bar{E}_{j}{\rm i}\gamma^{\mu}D_{\mu}E_{j}$ the
couplings $\tilde{g}_{Y}\frac{Y_{s}}{2} \sum_{j=0}^{N}\bar{E}_{j}
\gamma^{\mu}A_{j\mu}E_{j}$. In the mass eigenbasis, we have for the
right-handed fields
\begin{equation}
\mathcal{L}_{{ffA}} = \tilde{g}_{Y}\frac{Y_{s}}{2}\sum_{n,m,l}
\sum_{j}a_{jn}a_{jm}a_{jl}
\bar{\tilde{E}}_{nR}\tilde{A}_{m\mu}\gamma^{\mu}\tilde{E}_{lR}.
\end{equation}
Again using the orthogonality relations, we have
\begin{equation}
\mathcal{L}_{ffA} = -\tilde{g}_{Y}\frac{Y_{s}}{2\sqrt{N+1}}
\bar{\tilde{E}}_{1R}\tilde{A}_{1\mu}\gamma^{\mu}\tilde{E}_{0R} + {\rm
h.c.} + \ldots,
\end{equation}
where we have redefined the singlet fields $\tilde{E}_{n}\rightarrow
-\gamma_{5}\tilde{E}_{n}$, for $n=1,2,\ldots,N$, in order to obtain
positive masses in the singlet sector.

In the continuum model, the corresponding coupling in the singlet
sector is \cite{Bergstrom:2004nr,Appelquist:2000nn}
\begin{equation}\label{eq:conts}
\mathcal{L}_{{\rm cont.}} = -\frac{1}{4} g_{Y} Y_{s} \bar{E}_{R}^{(1)}
A_{\mu}^{(1)}\gamma^{\mu}(1+\gamma_{5})E^{(0)}+{\rm h.c.}
\end{equation}
Thus, with the identification $\tilde{g}_{Y}/\sqrt{N+1}=g_{Y}$, we have
the same coupling as for the continuum model.

\section{Indirect Detection of Kaluza--Klein Dark Matter}\label{sec:indet}

In models of continuous universal extra dimensions, the annihilation
of Kaluza--Klein gauge bosons can lead to an excess of positrons,
which could be observed in balloon-borne and space-based experiments.
The upcoming space-based experiments PAMELA and AMS-02 will reach
considerably higher energies than the balloon-borne HEAT experiment
did. The precision will also be largely improved. These improvements
are due to the large acceptance and long exposure time of the new
experiments. We focus in this paper on the PAMELA and AMS-02
experiments. 

The space-based (or more specifically satellite-borne) PAMELA
experiment is planned to be launched in June 2006. It
has an acceptance of $20~{\rm cm}^{2} \, {\rm sr}$ and an exposure
time of three years. The PAMELA experiment is optimized to measure the
cosmic positron spectrum for positron energies up to 270~GeV.

The space-based AMS-02 experiment will be placed on the International
Space Station and is planned to be launched in 2008. It has an
acceptance of $450~{\rm cm}^{2} \, {\rm sr}$ and an exposure time of
three years. The AMS-02 experiment is optimized to probe positron
energies up to 400~GeV.

In this section, we consider the differential positron flux from
Kaluza-Klein dark matter annihilations for the latticized model
described in the previous sections. We examine the latticization
effect and the prospects for upcoming experiments to probe this
effect. It should be noted that Kaluza--Klein dark matter could also
be observed indirectly by observing high-energy neutrinos or photons
from dark matter annihilations or directly from scattering off
nucleons \cite{Cheng:2002ej,Bergstrom:2004cy}. We leave the
consideration of these cases for future work.

\subsection{Annihilation of Gauge Bosons}\label{sec:annih}

\begin{figure}
\begin{center}
\includegraphics*{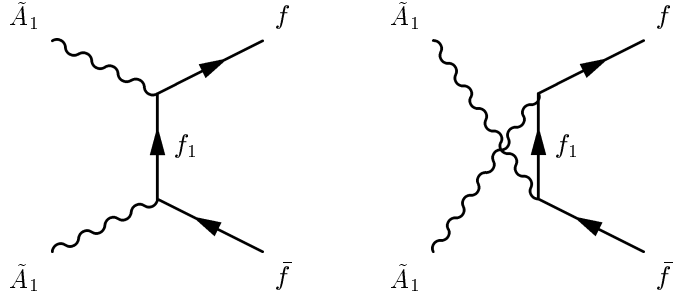}
\end{center}
\vspace*{-2mm}
\caption{\small{Kaluza--Klein gauge boson annihilation into SM fermion
pairs. Shown in the figure are the $t$- (left) and $u$-channel (right)
diagrams.}}\label{fig:annih}
\end{figure}

Positrons can be produced from several annihilation channels. They can
be produced both directly and indirectly through secondary
decays. Here we will follow \Ref~\cite{Cheng:2002ej} and only consider
positrons coming from direct $e^{+}e^{-}$ production. Positrons can
also be produced from cascade decays of muons, taus, and heavy quarks.
The lowest-order diagrams for the annihilation of the $U(1)$ gauge
bosons to particle-antiparticle pairs are given in
\Fig~\ref{fig:annih}, where we will consider $f=e$, \ie, direct
$e^{+}e^{-}$ production. Note that the fermionic propagators in
\Fig~\ref{fig:annih} can be both left- and right-handed.

The analysis will be almost identical to that of the continuum case,
we only have to replace the continuum Kaluza--Klein mode masses with
the corresponding masses of the latticized model. Thus, from the
Feynman rules in \Fig~\ref{fig:verticessimple} in
\App~\ref{app:Feynman}, we obtain for the $t$-channel annihilation
diagrams the amplitudes
\begin{equation}
\mathcal{M}_{L,R} = -{\rm i}g_{L,R}^{2}\bar{u} \gamma^{\mu}P_{L,R}
\bigg\lbrack \frac{\left(k_{\rho} -
p_{\rho}\right)\gamma^{\rho}}{(k-p)^{2}-m_{1}^{2}} \bigg\rbrack
\gamma^{\nu}P_{L,R}v\epsilon_{\mu}\epsilon_{\nu}.
\end{equation}
Here $u$ and $v$ denote four component spinors, $\epsilon_{\mu}$ and
$\epsilon_{\nu}$ denote polarization vectors, and the parameters
$g_{L,R}$ are given by
\begin{equation}
g_{L,R}=\frac{\tilde{g}_{Y}Y_{d,s}}{2\sqrt{N+1}}=\frac{g_{Y}Y_{d,s}}{2}.
\end{equation}
For the $u$-channel diagrams, one obtains similar expressions. The
cross-section for the continuum process has been calculated in
\Ref~\cite{Servant:2002aq}. In analogy with their result, we have
\begin{equation}
\sigma_{e^{+}e^{-}}=(g_{L}^{4}+g_{R}^{4})\frac{10(2m_{1}^{2}+s)\,{\rm
artanh}(\beta)-7s\beta}{72\pi s^{2}\beta^{2}},
\end{equation}
where
\begin{equation}
\beta=\sqrt{1-\frac{4m_{1}^{2}}{s}},
\end{equation}
$m_{1}=2M_{f}\sin\left\{\pi/[2(N+1)]\right\}$ is the mass in the
propagator, and $s=E_{CM}^{2}$ is a Mandelstam variable.

The differential positron flux is given by
\cite{Moskalenko:1999sb,Feng:2000zu}
\begin{equation}\label{eq:flux}
\frac{{\rm d}\Phi_{e^{+}}}{{\rm d}\Omega {\rm d}E} =
\frac{\rho_{0}^{2}}{m_{\tilde{A}_{1}}^{2}}
\sum_{i}\sigma_{i}vB_{e^{+}}^{i}\int {\rm d}\epsilon
f_{i}(\epsilon)G(\epsilon,E),
\end{equation}
where the sum is over all annihilation channels, $\rho_{0}$ is the
local dark matter density, and the $e^{+}$ branching fraction in
channel $i$ is denoted by $B_{e^{+}}^{i}$. The mass
$m_{\tilde{A}_{1}}=m_{1}$ is given by \eq~(\ref{eq:gbmass}). The Green
function $G(\epsilon,E)$ describes the propagation of the positrons
through the galaxy and $f_{i}(\epsilon)$ denotes the initial positron
energy distribution from channel $i$. As mentioned above, we will only
consider the flux of positrons coming from direct $e^{+}e^{-}$
production. Thus, we will only be concerned with the cross section
$\sigma_{e^{+}e^{-}}$. Since we are concerned with non-relativistic
particles, we can make a series expansion of $\sigma_{e^{+}e^{-}} v$
for small velocities $v$. Thus, we obtain as in
\Ref~\cite{Cheng:2002ej}
\begin{equation}\label{eq:sigmav}
\sigma_{e^{+}e^{-}} v \simeq \frac{\tilde{g}_{Y}^{4}}{288\pi
(N+1)^{2}m_{1}^{2}}\left(Y_{d}^{4}+Y_{s}^{4}\right).
\end{equation}
Here we have written \eq~(\ref{eq:sigmav}) explicitly in terms of the
parameters of our original model. Recall that we have made the
identification $\tilde{g}_{Y}/\sqrt{N+1}=g_{Y}$. Note also that this
result includes a factor of 1/16, which is not present in
\Ref~\cite{Cheng:2002ej}. This is because we have a different
convention for the hypercharge assignment.

In calculating the differential positron flux, we have used the {\sc
DarkSUSY} package, see \Refs~\cite{Gondolo:2000ee,Gondolo:2004sc}. In
\Figs~\ref{fig:plot300}-\ref{fig:plot900}, we present the differential
positron flux as a function of the positron energy for an inverse
radius of 300~GeV, 450~GeV, 600~GeV, 750~GeV and 900~GeV
respectively. We present the results for latticized models with $N=1$
(\ie, two lattice sites), $N=2$, and $N=3$. In addition, we give the
continuum model results. We have chosen an isothermal sphere dark
matter distribution\footnote{The Frenk--Navarro--White (FNW)
distribution gives very similar results.}  with a halo size of $z_{h}=
4\ {\rm kpc}$ and a local dark matter density of $\rho_{0}=0.3~{\rm
GeV} \, {\rm cm}^{-3}$. The propagation parameters are taken from
\Ref~\cite{Edsjo:2004pf}. In
\Figs~\ref{fig:plot300}-\ref{fig:plot900}, we also make a comparison
with a model for the background differential positron flux. In all
figures, we have chosen ``model C'' from \Ref~\cite{Strong:1998fr}. We
have included a factor 1/2 when calculating the differential positron
flux, which comes from the fact that the gauge bosons always
annihilate in pairs \cite{Bringmann:2005pp}. This factor has not
always been accounted for in previous works, which is why our
continuum results may differ by a factor 1/2 from some previous
results on the continuum positron spectrum.

In \Ref~\cite{Oliver:2003cc}, the bounds from electroweak precision
observables on the mass $m_{1}$ of the first Kaluza--Klein mode for
latticized and continuous universal extra dimensions are
discussed. Note that for a continuous extra dimension $m_{1}$ equals
the inverse radius $R^{-1}$, whereas for a latticized model we have
that $m_{1}= 2(N+1)(\pi R)^{-1}\sin\left\{\pi/[2(N+1)]\right\}$. In
\Ref~\cite{Oliver:2003cc}, the most stringent bound for a single
continuous universal dimension was found to be $m_{1}=R^{-1}\gtrsim
400$~GeV ($95~\%$ CL). It was also found in \Ref~\cite{Oliver:2003cc}
that the bound for a few-site lattice model can be lowered by
10~\%-25~\%, which would allow for a bound as low as $m_{1}\gtrsim
300$~GeV ($95~\%$ CL) for a few-site lattice model. The reason why the
bounds are less restrictive for the lattice model is the realization
of only a few Kaluza--Klein modes. This general feature of lattice
models could be important for the PAMELA and AMS-02 experiments, which
for a continuum model could be out of range of the acceptable
energies. In a recent paper, \Ref~\cite{Flacke:2005hb}, the authors
argue that the bounds from electroweak precision observables could be
improved, giving a bound as severe as $m_{1} = R^{-1}\gtrsim 700$~GeV
($99$~\% CL). This bound was obtained by taking into account two-loop
effects and LEP2 data. However, as argued above, the bounds for a
lattice model should be less severe, especially for a few-site lattice
model. There are also bounds from WMAP for the relic density of
Kaluza--Klein dark matter \cite{Bennett:2003bz}. It was found in
\Refs~\cite{Burnell:2005hm,Kong:2005hn} that in the minimimal
continuum UED model, including coannihilations, to account for the
observed relic density, the ideal mass range for the lightest
Kaluza--Klein particle should be $500$~GeV$- 600$~GeV. Note that the
bounds from WMAP can be lowered by assuming that Kaluza--Klein dark
matter only makes up a fraction of the cold dark matter in the
Universe. Of course, in this case, the annihilation rate in the halo
will be reduced, making the detection of positrons with PAMELA and
AMS-02 more difficult. In addition, there are bounds from direct
detection \cite{Servant:2002hb}, where the most stringent bound for a
continuum model gives $m_{1}\gtrsim 400$~GeV \cite{Sanglard:2005we},
which is no more restrictive than the bounds from electroweak
precision observables.

Note the peak in the positron spectrum at a positron energy equal to
the WIMP mass (\ie, the mass of the first excited mode of the $U(1)$
gauge boson, $E_{e^{+}}=m_{\tilde{A}_{1}}=m_{1}$). This peak is due to
the monoenergetic positron source and is a characteristic signature of
Kaluza--Klein dark matter. In contrast, for neutralinos, the positron
spectrum would be much smoother \cite{Ellis:2001hv}. This is due to
the Majorana nature of neutralinos, which leads to helicity
suppression of direct $e^+ e^-$ production. Note that since the
lattice model only differ from the continuum model in the masses of
the Kaluza--Klein modes, one could not by measuring the differential
positron flux distinguish between a lattice model with a given radius
and a continuum model with a larger radius, \ie, both models would
make the same prediction.\footnote{At least this conclusion holds at
tree-level. We will not investigate radiative corrections.} Such a
degeneracy would also exist between a lattice model with parameters
$N_{1}$ and $R_{1}$ and a different lattice model with parameters
$N_{2}$ and $R_{2}$. Thus, one would require additional input from
independent, different types of experiments\footnote{For example, one
could compare the level-spacing in the mass spectrum between the
zeroth and the first Kaluza--Klein modes to the spacing between the
first and the second modes. For the continuum model, the spacing will
be the same, whereas for the lattice model the spacing will decrease.
Such an experiment would also uniquely determine $N$ and $R$.} in
order to use the information from the differential positron flux to
probe lattice effects.

We also observe that the lattice model converges very quickly to the
continuum model. Given the uncertainties in astrophysical inputs and
propagation models, it could be hard to detect such small
deviations. For increasing values of $N$, the magnitude (\ie, the
height of the peak) of the differential positron flux decreases and
the peak is shifted towards the peak of the continuum model, whereas
for decreasing values of the radius $R$ of the extra dimension, the
separation among the peaks for different values of $N$ increases and
the differential positron flux (as well as the height of the peak)
decreases. This can be observed in \eq~({\ref{eq:flux}), which shows
that the decrease comes from the fact that the differential positron
flux is suppressed by the mass of the Kaluza--Klein mode. Since the
mass of the Kaluza--Klein mode is inversely proportional to the radius
of the extra dimension, the flux will be smaller for a smaller radius,
which can also be seen when comparing
\Figs~\ref{fig:plot300}-\ref{fig:plot900}.
\begin{figure}
\begin{center}
\includegraphics*[scale=0.46]{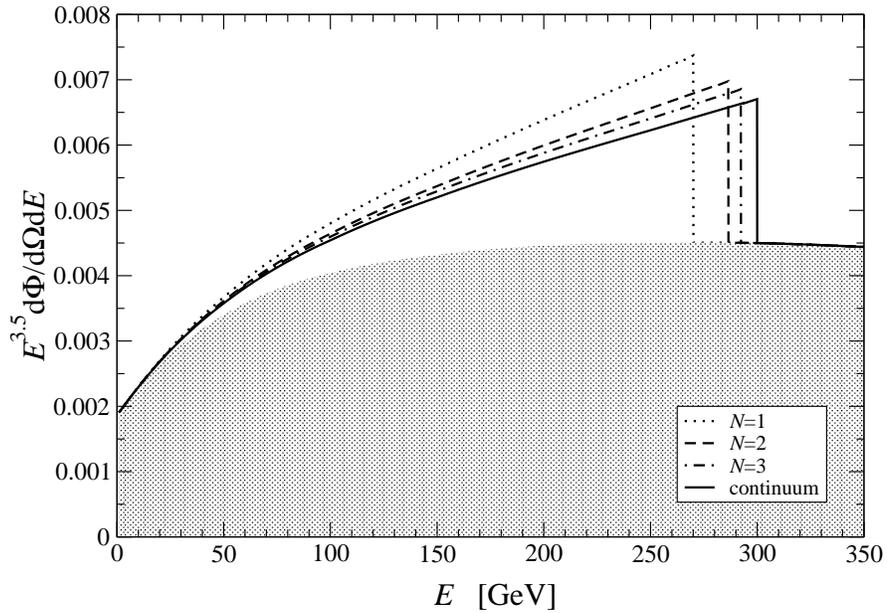}
\end{center}
\vspace*{-2mm}
\caption{\small{The differential positron flux (above background) for
an inverse radius of 300~GeV as a function of positron energy, where
we have only considered direct $e^{+}e^{-}$ production. In addition,
positrons can also be produced from cascade decays of muons, taus and
heavy quarks. Presented are latticized models with two lattice sites
($N=1$, dotted curve), three lattice sites ($N=2$, dashed curve) and
four lattice sites ($N=3$, dash-dotted curve) as well as the continuum
model (solid curve). Given is also an estimated background flux (gray
shaded). The unit of the ordinate is
cm$^{-2}$s$^{-1}$sr$^{-1}$GeV$^{2.5}$.}}\label{fig:plot300}
\end{figure}

\begin{figure}
\begin{center}
\includegraphics*[scale=0.46]{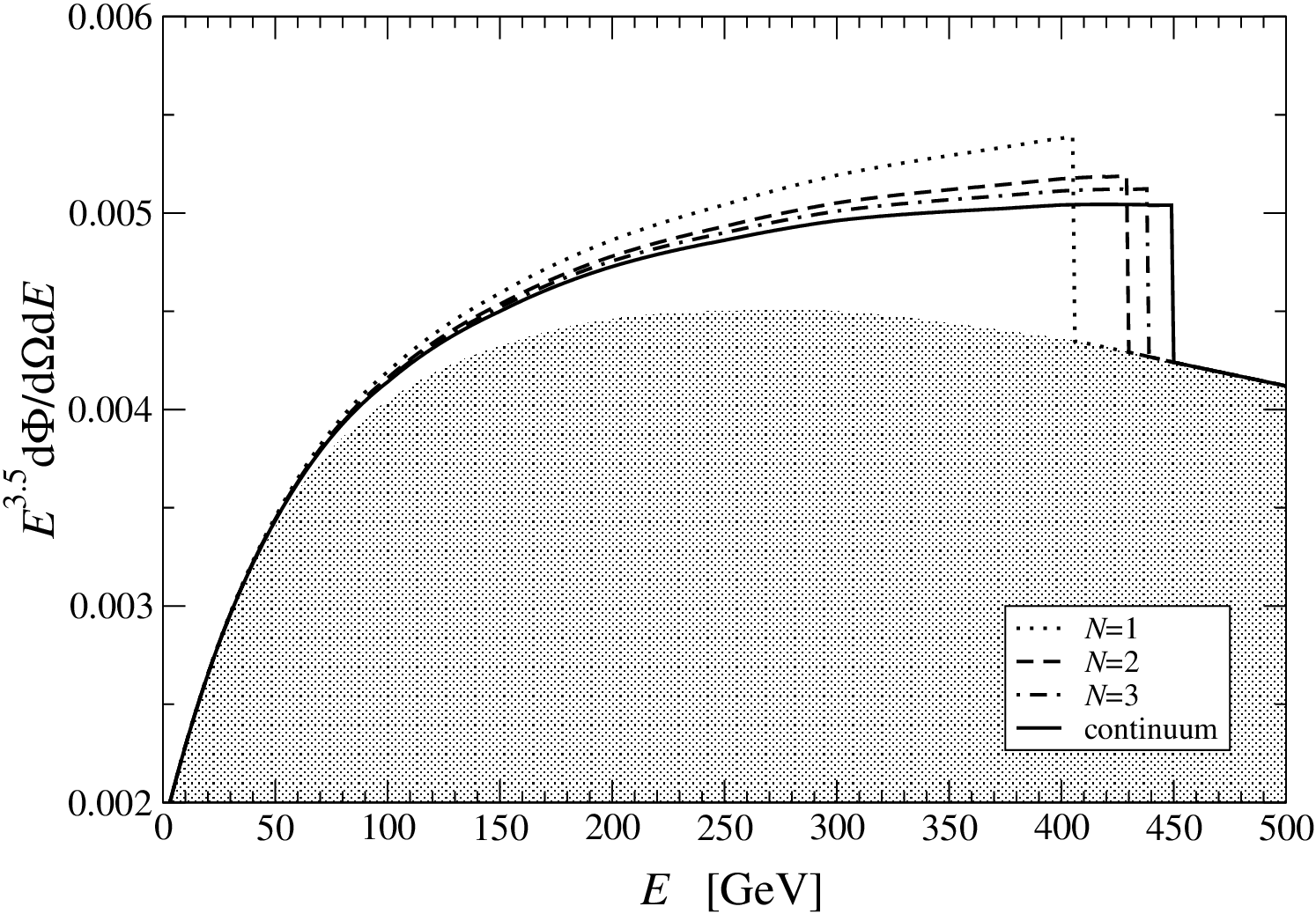}
\end{center}
\vspace*{-2mm}
\caption{\small{The differential positron flux (above background) for
an inverse radius of 450~GeV as a function of positron energy, where
we have only considered direct $e^{+}e^{-}$ production. In addition,
positrons can also be produced from cascade decays of muons, taus and
heavy quarks. Presented are latticized models with two lattice sites
($N=1$, dotted curve), three lattice sites ($N=2$, dashed curve) and
four lattice sites ($N=3$, dash-dotted curve) as well as the continuum
model (solid curve). Given is also an estimated background flux (gray
shaded). The unit of the ordinate is
cm$^{-2}$s$^{-1}$sr$^{-1}$GeV$^{2.5}$. For presentation purposes we
have chosen the ordinate to start at $0.002$.}}\label{fig:plot450}
\end{figure}

\begin{figure}
\begin{center}
\includegraphics*[scale=0.46]{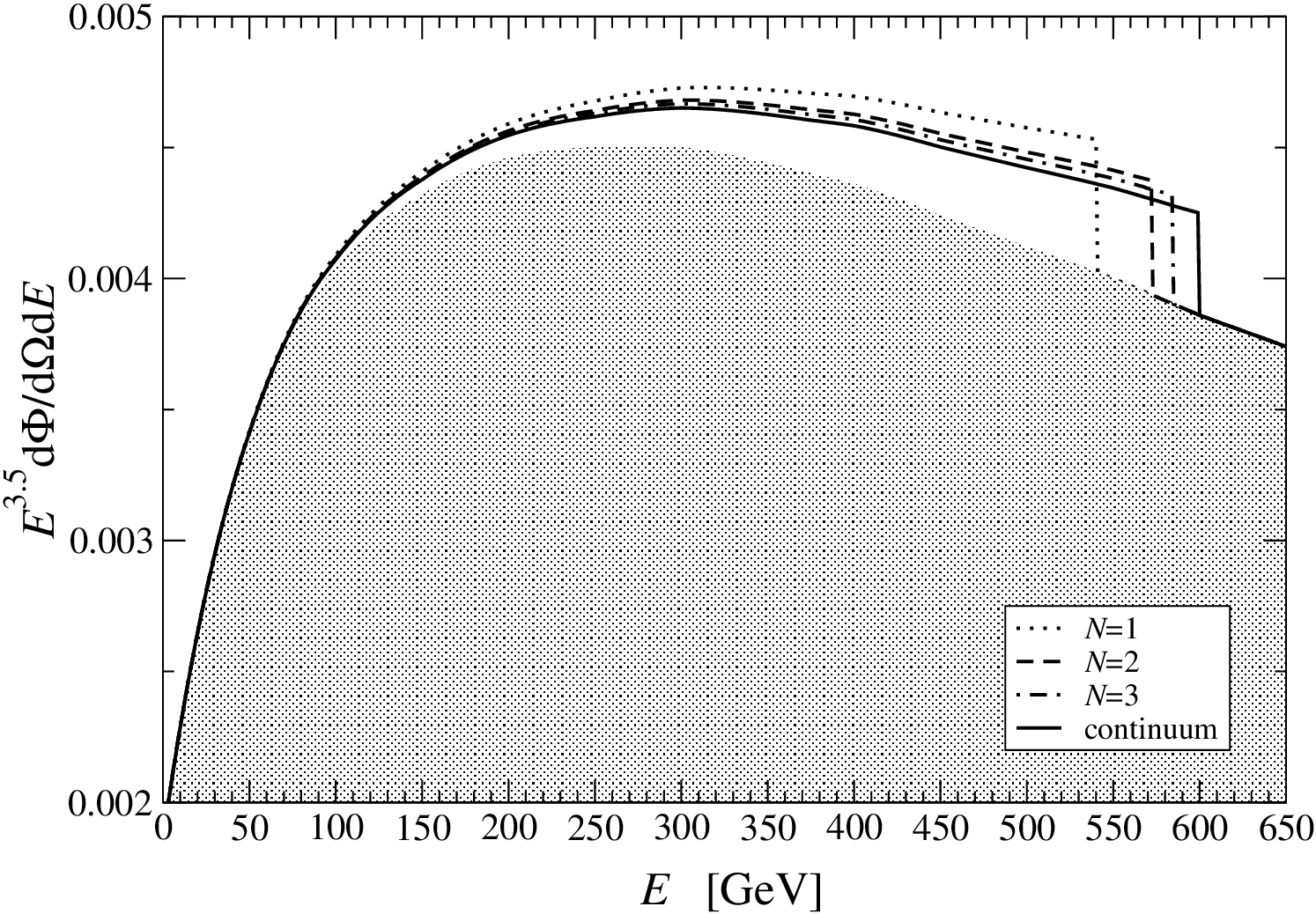}
\end{center}
\vspace*{-2mm}
\caption{\small{The differential positron flux (above background) for
an inverse radius of 600~GeV as a function of positron energy, where
we have only considered direct $e^{+}e^{-}$ production. In addition,
positrons can also be produced from cascade decays of muons, taus and
heavy quarks. Presented are latticized models with two lattice sites
($N=1$, dotted curve), three lattice sites ($N=2$, dashed curve) and
four lattice sites ($N=3$, dash-dotted curve) as well as the continuum
model (solid curve). Given is also an estimated background flux (gray
shaded). The unit of the ordinate is
cm$^{-2}$s$^{-1}$sr$^{-1}$GeV$^{2.5}$. For presentation purposes we
have chosen the ordinate to start at $0.002$.}}\label{fig:plot600}
\end{figure}

\begin{figure}
\begin{center}
\includegraphics*[scale=0.46]{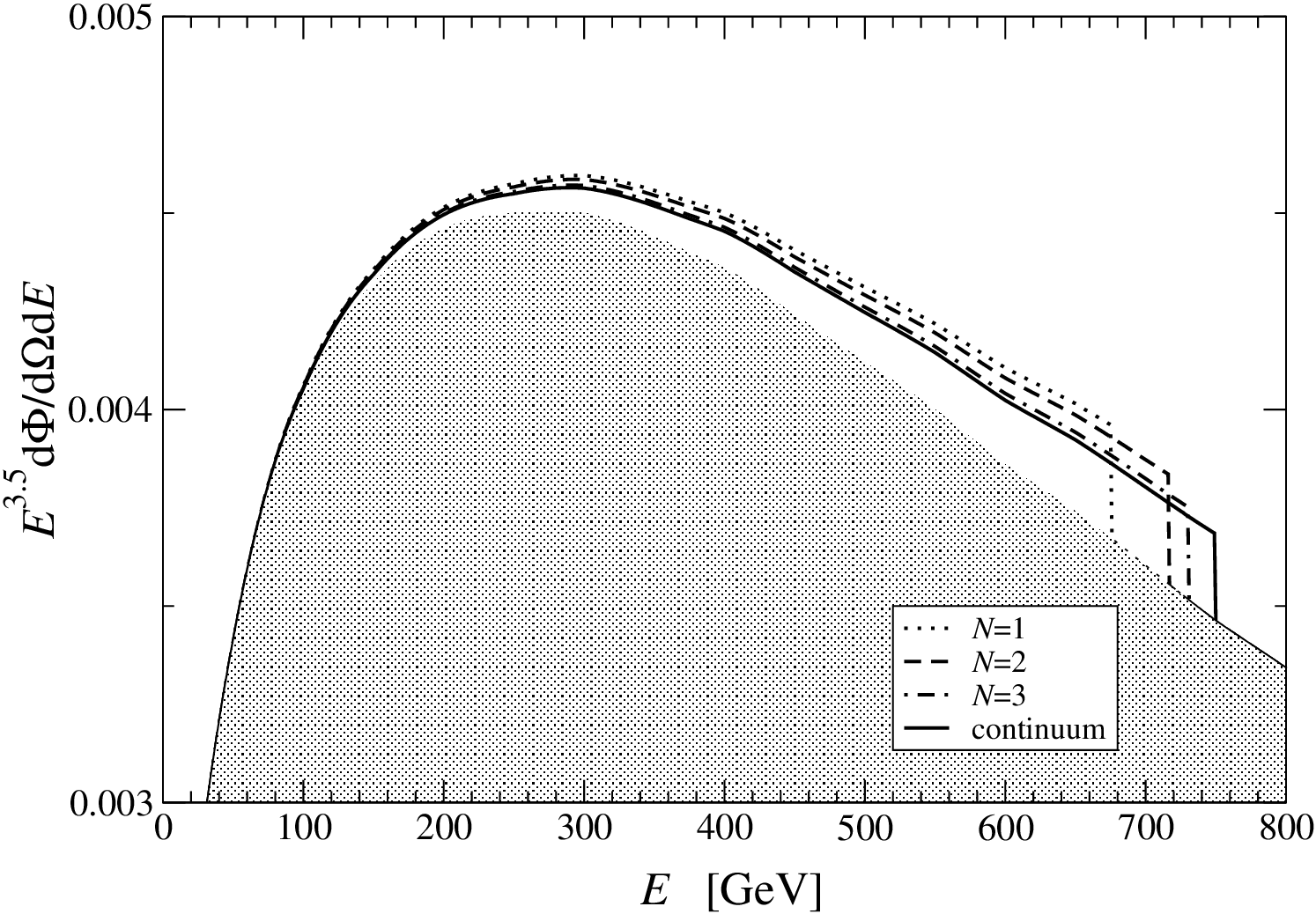}
\end{center}
\vspace*{-2mm}
\caption{\small{The differential positron flux (above background) for
an inverse radius of 750~GeV as a function of positron energy, where
we have only considered direct $e^{+}e^{-}$ production. In addition,
positrons can also be produced from cascade decays of muons, taus and
heavy quarks. Presented are latticized models with two lattice sites
($N=1$, dotted curve), three lattice sites ($N=2$, dashed curve) and
four lattice sites ($N=3$, dash-dotted curve) as well as the continuum
model (solid curve). Given is also an estimated background flux (gray
shaded). The unit of the ordinate is
cm$^{-2}$s$^{-1}$sr$^{-1}$GeV$^{2.5}$. For presentation purposes we
have chosen the ordinate to start at $0.003$.}}\label{fig:plot750}
\end{figure}

\begin{figure}
\begin{center}
\includegraphics*[scale=0.46]{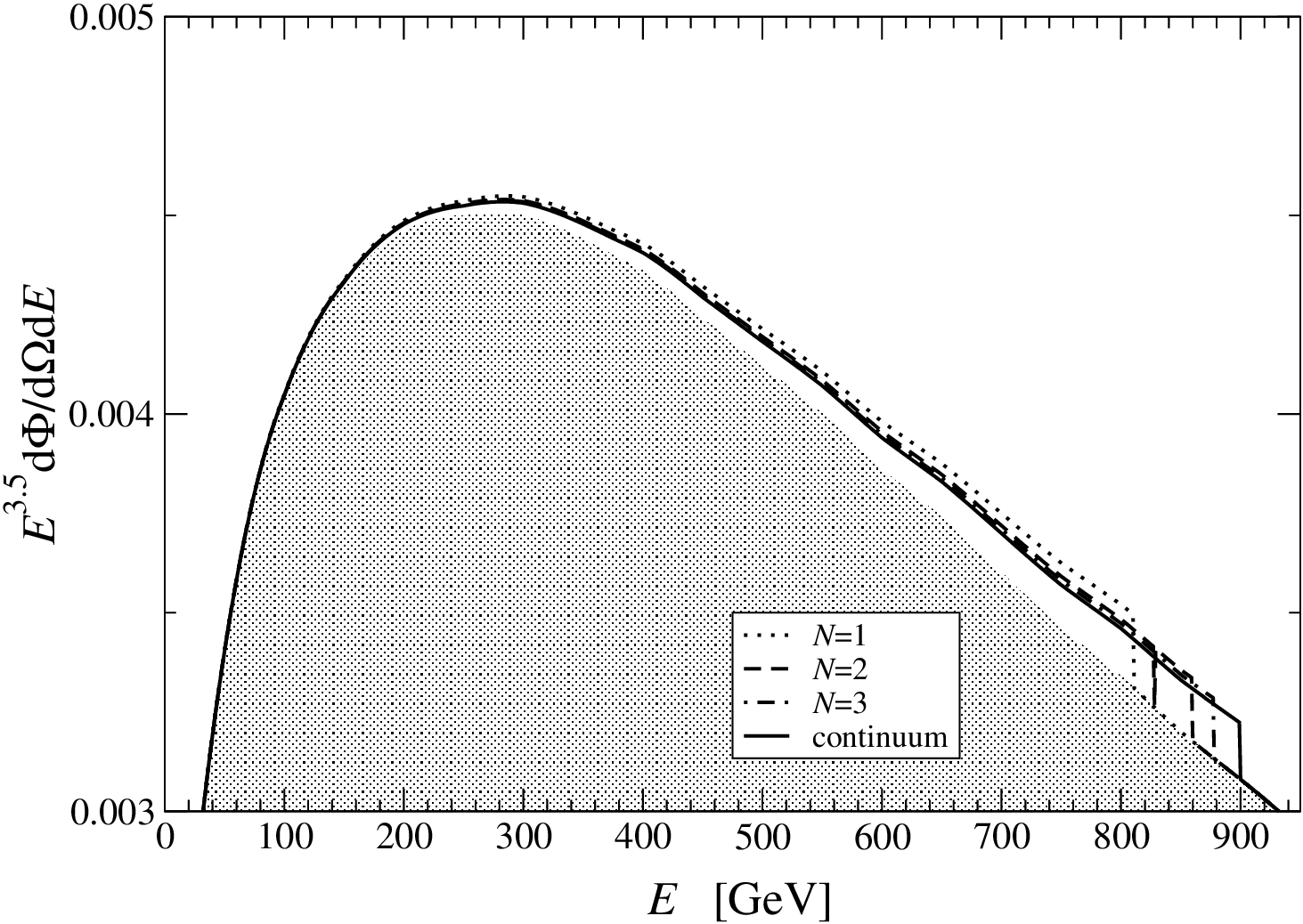}
\end{center}
\vspace*{-2mm}
\caption{\small{The differential positron flux (above background) for
an inverse radius of 900~GeV as a function of positron energy, where
we have only considered direct $e^{+}e^{-}$ production. In addition,
positrons can also be produced from cascade decays of muons, taus and
heavy quarks. Presented are latticized models with two lattice sites
($N=1$, dotted curve), three lattice sites ($N=2$, dashed curve) and
four lattice sites ($N=3$, dash-dotted curve) as well as the continuum
model (solid curve). Given is also an estimated background flux (gray
shaded). The unit of the ordinate is
cm$^{-2}$s$^{-1}$sr$^{-1}$GeV$^{2.5}$. For presentation purposes we
have chosen the ordinate to start at $0.003$.}}\label{fig:plot900}
\end{figure}

\section{Modified Boundary Conditions}\label{sec:MBC}

Finally, we will consider a model with more complicated boundary
conditions. For the gauge fields, we impose that
\begin{equation}
A_{N}-A_{N-1}=0,
\end{equation}
which is a discretized version of the continuum Neumann boundary
condition. This will lead to the same type of gauge boson mass matrix
as in \Sec~\ref{sec:LUD}, but now it will be an $N\times N$ matrix
instead of an $(N+1) \times (N+1)$ matrix. Thus, the expansion in mass
eigenstates will now take the form
\begin{equation}
A_{j}=\sum_{n=0}^{N-1}a_{jn}\tilde{A}_{n},
\end{equation}
where the $a_{jn}$'s are defined as in \eq~(\ref{eq:ajn}), but with
the replacement $N+1 \rightarrow N$. The mass eigenvalues will now be
$m_{n}^{2} = \tilde{g}_{Y}^{2} v_{1}^{2} Y_{\phi}^{2} \sin^{2}
\left[n\pi/(2N)\right]$, where $n = 0,1,\ldots N-1$. The
correspondence with the continuum mass spectrum is obtained by
requiring that $\pi \tilde{g}_{Y} v_{1} Y_{\phi} /(2N) = 1/R$.

We impose as in \Ref~\cite{Cheng:2001vd} discretized Neumann boundary
conditions in the fermionic sector by requiring that $L_{jL}$ and
$E_{jR}$ satisfy
\begin{equation}
L_{NL} - L_{N-1,L}=0 \quad \mbox{and} \quad E_{NR} - E_{N-1,R}=0.
\end{equation}
We take the $L_{jR}$ and $E_{jL}$ fields to satisfy discretized
Dirichlet boundary conditions, \ie, we impose that $L_{0R}=L_{NR}=0$
and $E_{0L}=E_{NL}=0$. In this way, we obtain, as in the continuum
model, chiral zeroth modes. The corresponding mass matrix is
diagonalized by a change of basis
\begin{equation}
L_{jL}=\sum_{n=0}^{N-1}a_{jn}\tilde{L}_{nL}\,,\quad\quad
L_{jR}=\sum_{n=1}^{N-1}b_{jn}\tilde{L}_{nR}\,,
\end{equation}
where the $a_{jn}$'s and $b_{jn}$'s are given in \eqs~(\ref{eq:ajn})
and (\ref{eq:bjn}), respectively, again with the replacement
$N+1\rightarrow N$. We obtain for the left- and right-handed handed
fields the masses $m_n^2 = 4 M_f^{2} \sin^2
\left[n\pi/(2N)\right]$. For $n\ll N$, we find a linear Kaluza--Klein
spectrum by making the identification $\pi M_{f}/N=1/R$.

Similarly, we obtain a mass matrix for the singlet sector, which is
diagonalized by a change of basis
\begin{equation}
E_{jL}=\sum_{n=1}^{N-1}b_{jn}\tilde{E}_{nL}\,, \quad\quad
E_{jR}=\sum_{n=0}^{N-1}a_{jn}\tilde{E}_{nR}.
\end{equation}
We obtain as in \Sec~\ref{sec:fermions} negative masses for the
singlet fields, which can be remedied by a change of basis
$\tilde{E}_{n}\rightarrow -\gamma_{5}\tilde{E}_{n}$, for
$n=1,2,\ldots, N-1$.

\subsection{Fermion Gauge Boson Couplings}

Analogously to \Sec~\ref{sec:ffA}, we derive in this section the
couplings between fermion and gauge bosons for the model with modified
boundary conditions. First, we consider the $SU(2)$ doublet fields and
find for the left-handed fields the couplings
\begin{eqnarray}\label{eq:dcoupl}
\mathcal{L}_{ffA} &=& \tilde{g}_{Y}\frac{Y_{d}}{2}
\bigg\{\Delta_{0}\tilde{A}_{1\mu}\bar{\tilde{L}}_{0L}
\gamma^{\mu}\tilde{L}_{0L} + \big\lbrack(
\Delta_{1}+\sqrt{1/N})\tilde{A}_{1\mu}\bar{\tilde{L}}_{0L}
\gamma^{\mu}\tilde{L}_{1L} + {\rm h.c.} \big\rbrack\nonumber\\ &&+
\sum_{l=2}^{N-1} \left( \Delta_{l}\tilde{A}_{1\mu}\bar{\tilde{L}}_{0L}
\gamma^{\mu}\tilde{L}_{lL} + {\rm h.c.} \right) + \ldots \bigg\},
\end{eqnarray}
where again the dots indicate terms that are not directly relevant to
our discussion and
\begin{equation}\label{eq:deltak}
\Delta_{k} = \frac{1}{\sqrt{N}}a_{N1}a_{Nk} = \left\{
\begin{array}{ll} \frac{2}{\sqrt{N^{3}}}
\cos\left(\frac{2N+1}{2}\gamma_{1}\right)\,
\cos\left(\frac{2N+1}{2}\gamma_{k}\right) , & k\neq 0\\\\
\sqrt{\frac{2}{N^{3}}}\cos\left(\frac{2N+1}{2}\gamma_{1}\right), & k=0
\end{array} \right..
\end{equation}
For small or moderate $N$, the couplings in \eq~(\ref{eq:dcoupl})
deviate from the continuum results, see \eq~(\ref{eq:contd}). The
deviation is encoded in the parameters $\Delta_{k}$. Each of them goes
as $1/N$, after one factor $1/\sqrt{N}$ has been absorbed in the
redefinition of the coupling constant. Thus, for large number of
lattice sites, the contribution from the $\Delta_{k}$ parameters
become negligible and we reproduce the continuum results.

Second, in the singlet sector, we find the following couplings 
\begin{eqnarray}\label{eq:scoupl}
\mathcal{L}_{ffA} &=& \tilde{g}_{Y}\frac{Y_{s}}{2}\bigg\{\Delta_{0}
\tilde{A}_{1\mu}\bar{\tilde{E}}_{0R} \gamma^{\mu}\tilde{E}_{0R} -
\big\lbrack(\Delta_{1}+\sqrt{1/N})\tilde{A}_{1\mu}\bar{\tilde{E}}_{0R}
\gamma^{\mu}\tilde{E}_{1R} + {\rm h.c.} \big\rbrack\nonumber\\ &&-
\sum_{l=2}^{N-1} \left(\Delta_{l}\tilde{A}_{1\mu}\bar{\tilde{E}}_{0R}
\gamma^{\mu}\tilde{E}_{lR} + {\rm h.c.} \right) + \ldots\bigg\},
\end{eqnarray}
where again we have redefined the singlet fields
$\tilde{E}_{n}\rightarrow -\gamma_{5}\tilde{E}_{n}$, for
$n=1,2,\ldots, N-1$, in order to obtain positive masses in the singlet
sector. Precisely as for the doublet sector, the couplings in
\eq~(\ref{eq:scoupl}) deviate for small $N$ from the corresponding
continuum couplings, see \eq~(\ref{eq:conts}). However, in the large
$N$ limit, we recover the continuum results.

Thus, we observe that for a small or moderate number of lattice sites,
we obtain for the singlet and doublet sectors additional vertices,
which are not present in the continuum. The Feynman rules for this
model are given in \Figs~\ref{fig:verticesd} and \ref{fig:verticess}
in \App~\ref{app:Feynman}. The new Feynman rules will lead to
additional diagrams for the annihilation process considered in
\Sec~\ref{sec:annih}. Thus, the positron flux will have a more
complicated dependence on the latticization than in the model with
the simpler boundary conditions, \cf, \Sec~\ref{sec:LUD}. In particular,
the degeneracy between the lattice model and a continuum model with
larger radius is broken. However, we also observe that the analogue of
Kaluza--Klein parity is explicitly broken. Note that Kaluza--Klein
parity is necessary to ensure the stability of the lightest
Kaluza--Klein mode, which is necessary for it to be a viable dark
matter candidate. As the number of lattice sites is increased,
Kaluza--Klein parity is approximately conserved. However, as we have
seen, the lattice model converges very quickly to the continuum
results and so in the region were Kaluza--Klein parity is
approximately conserved, the deviance may anyway be too small to
detect. Therefore, at this level, the model considered in this section
is mostly of academic interest. A deeper analysis would be required to
determine the phenomenological relevance of this model, but such an
analysis is beyond the scope of this paper.

\section{Summary and Conclusions}
\label{sec:S&C}

We have considered Kaluza--Klein dark matter from latticized universal
dimensions. We have studied two different models for latticized
universal dimensions, where the models differ in the choice of
boundary conditions. We have examined to what extent the models can
reproduce the continuum results for Kaluza--Klein dark matter. For the
model with simple boundary conditions, we have found that we can
reproduce the essential features of the continuum model, such as
chiral zeroth modes and the relevant couplings, even for a few-site
lattice model. We have especially examined the effects of the
latticization on the differential positron flux from Kaluza--Klein
dark matter annihilation and the prospects for upcoming experiments
such as PAMELA and AMS-02 to probe the latticization effect. For the
model with simple boundary conditions, we have found that the results
would be equivalent to a continuum model with a larger radius. Thus,
in conclusion, the results from such an experiment should be used in
conjunction with the result from some different independent experiment
in order to be able to probe lattice effects. We have also pointed out
that, since the experimental bounds on latticized universal dimensions
are less severe than for the corresponding continuum model, the
prospects for detection for the lattice model are better. This could
be important for the PAMELA and AMS-02 experiments, since these
experiments could be out of range to detect the results of the
continuum model.

For the model with modified boundary conditions, we have found a
non-trivial dependence on the latticization. For a small number of
lattice sites, Kaluza--Klein parity is violated, which means that, in
this range, the model may not be a viable model for dark matter. For a
large number of lattice sites, Kaluza--Klein parity is approximately
conserved. However, due to the fast convergence to the continuum
results, the prospects for detecting lattice effects in this range
will be small.

\subsection*{Acknowledgments}

We would like to thank Joakim Edsj{\"o} for help with {\sc DarkSUSY}
and Thomas Konstandin, Konstantin Matchev, Mark Pearce, and Gerhart
Seidl for useful discussions.

This work was supported by the Royal Swedish Academy of Sciences
(KVA), the Swedish Research Council (Vetenskapsr{\aa}det), Contract
Nos.~621-2001-1611, 621-2002-3577, the G{\"o}ran Gustafsson
Foundation, and the Magnus Bergvall Foundation.

\appendix

\section{Orthogonality Relations}
\label{app:orthogonal} 

In this appendix, we list some useful orthogonality relations for our
formalism. In all of these relations, we have assumed that $0\leqslant
n,m \leqslant N$. In the main text, when diagonalizing the mass
matrices, we consider in \eq~(\ref{eq:ajn}) the expansion coefficients
\begin{equation}
a_{jn}=\left\{ \begin{array}{ll} 
\sqrt{\frac{2}{N+1}}\cos\left(\frac{2j+1}{2}\gamma_{n}\right), & n\neq 0\\
\sqrt{\frac{1}{N+1}}, & n=0 \end{array} \right., 
\end{equation}
where $\gamma_{n}=n\pi/(N+1)$. In addition, we have from \eq~(\ref{eq:bjn})
the expansion coefficients
\begin{equation}
b_{jn}=\sqrt{\frac{2}{N+1}}\sin\left(j\gamma_{n}\right).
\end{equation}
Note that the expansion coefficients obey the following relations
\begin{equation}
\sum_{j=0}^{N}a_{jn}a_{jm}=\delta_{nm} \quad \mbox{and} \quad
\sum_{j=0}^{N}b_{jn}b_{jm}=\delta_{nm}.
\end{equation}
Thus, we have for $n,m,l\neq 0$ and $n,m,l\ll N$
\begin{equation}
\sum_{j=0}^{N}a_{jn}a_{jm}a_{jl} = \frac{1}{\sqrt{2(N+1)}}
\left(\delta_{l,n+m}+\delta_{m,n+l}+\delta_{n,m+l}\right).
\end{equation}
Note that if one of the indices $n$, $m$, and $l$ is zero, then we
obtain
\begin{equation}
\sum_{j=0}^{N}a_{jn}a_{jm}a_{jl} = \frac{1}{\sqrt{N+1}}
\left(\delta_{n0}\delta_{ml} + \delta_{m0}\delta_{nl} +
\delta_{l0}\delta_{nm}\right).
\end{equation}
Furthermore, if two are zero and one is non-zero, then we obtain
zero. Finally, if all three of the indices $n$, $m$, and $l$ are zero,
\ie, $(n,m,l)=(0,0,0)$, then we obtain
\begin{equation}
\sum_{j=0}^{N}a_{jn}a_{jm}a_{jl}=\frac{1}{\sqrt{N+1}}.
\end{equation}
The corresponding relations for the $b_{jn}$'s will not be relevant
for the purpose of this paper.

\section{Feynman Rules}
\label{app:Feynman}

In this appendix, we give the relevant Feynman rules for the models
considered in the text. The Feynman rules are derived by first going
to the mass eigenbasis and the simply reading off the rules for the
vertices. In \Fig~\ref{fig:verticessimple}, we give the Feynman rules
for the model considered in \Sec~\ref{sec:LUD}, and in
\Figs~\ref{fig:verticesd} and \ref{fig:verticess}, we give the Feynman
rules for the model with modified boundary conditions considered in
\Sec~\ref{sec:MBC}. Here $P_{R,L}=\frac{1}{2}(1\pm \gamma_{5})$ are
the chirality projectors.
\vspace{1cm}
\begin{figure}[!h]
\begin{center}
\includegraphics*[scale=1.0]{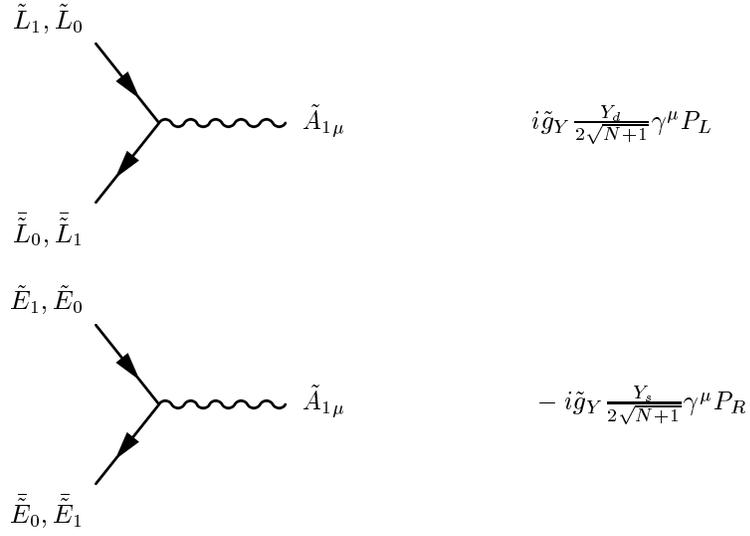}
\end{center}
\vspace*{0.5cm}
\caption{\small{Feynman rules for the model with simple boundary
conditions.}}
\label{fig:verticessimple}
\end{figure}

\begin{figure}[!h]
\begin{center}
\includegraphics*[scale=1.0]{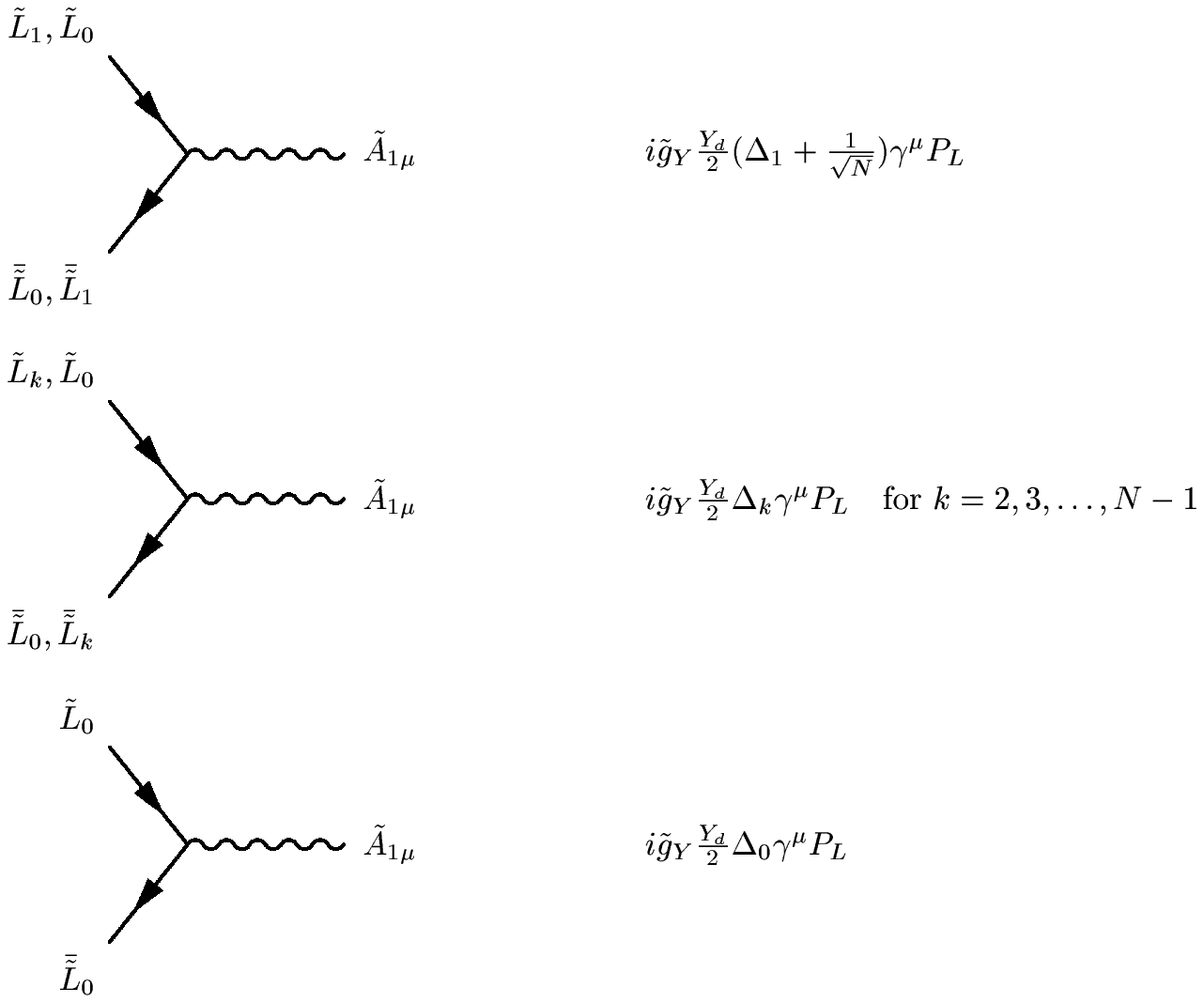}
\end{center}
\vspace*{0.5cm}
\caption{\small{Feynman rules for the doublet sector of the model with
modified boundary conditions.}}
\label{fig:verticesd}
\end{figure}

\begin{figure}[!h]
\begin{center}
\includegraphics*[scale=1.0]{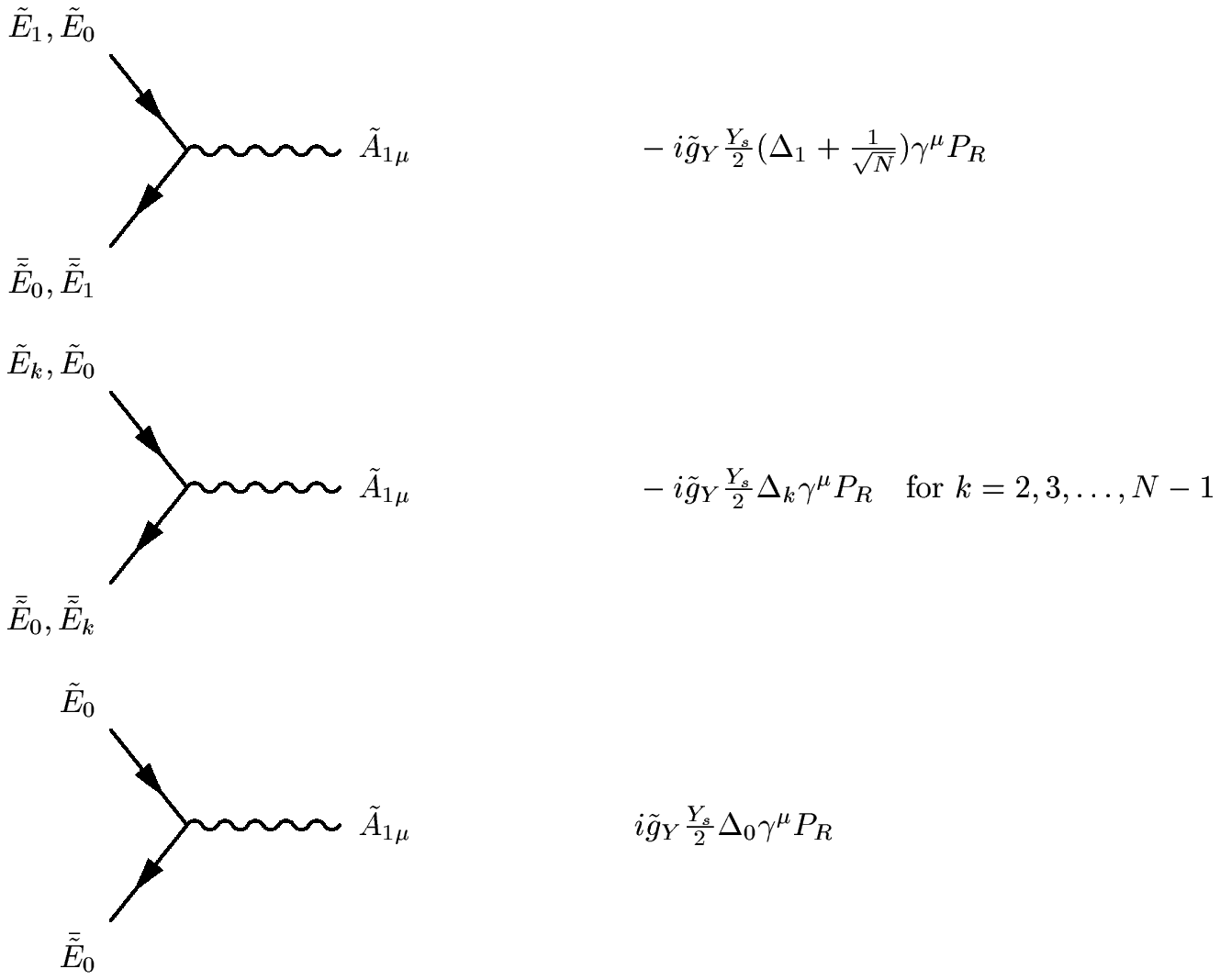}
\end{center}
\vspace*{0.5cm}
\caption{\small{Feynman rules for the singlet sector of the model with
modified boundary conditions.}}
\label{fig:verticess}
\end{figure}

\end{document}